\documentstyle[12pt,psfig,epsf]{article}
\newcommand{\be}{\begin{equation}}
\newcommand{\ee}{\end{equation}}

\input epsf.tex

\begin{document}
\pagestyle{plain}
\begin{titlepage}

\begin{flushright}
PUPT-1818\\
hep-th/9810163
\end{flushright}
\vspace{15 mm}

\begin{center}
{\huge Brane Death via Born-Infeld String  }
\end{center}

\vspace{12 mm}

\begin{center}
Konstantin G.Savvidy  \footnote{ksavvidi@princeton.edu} \\
\vspace{3mm}
Joseph Henry Laboratories\\
Princeton University\\
Princeton, New Jersey 08544
\end{center}

\vspace{5 mm}
\begin{abstract}
\noindent

I revisit the solution of Born-Infeld theory which
corresponds to a 3-brane and anti-brane joined by a
(fundamental) string. The global instability of this configuration
makes possible the semiclassical tunneling 
into a wide, short tube which keeps expanding
out, thus annihilating the brane . This tunneling is suppressed
exponentially as $\exp\{-\frac{S_{cl}}{g}\}$. The attraction between
the branes causes them to approach and at some point to tunnel, 
because the action of the bounce solution goes to zero.
The energy of the solution at the top of the barrier, the sphaleron, goes like
$\sim D^3$ for large separarions D, while the energy of the string is proportional
to its length D. 

\end{abstract}
\end{titlepage}
\pagestyle{plain}
\section{Introduction}
\vspace{.5cm}
 Recently, Callan and Maldacena \cite{callan} considered among other
configurations the 3-brane and anti-brane joined by a (fundamental) string
in the framework of Born-Infeld theory. The string is made of a 3-brane,
wrapped aroung $S^2$ sphere. When looked from some distance, such an object
does not appear to carry RR charge as a whole,but is rather like a RR dipole,
and has energy per unit length proper to the fundamental string. 

Such a configuration is only quasi-stable,
since globally it is possible to lose energy by making the throat very wide:
if it had radius R, the change in energy is mostly due to tension\footnote{
The second term has the same origin as in \cite{emparan}. The two
branes act as a sort of capacitor to create a uniform bulk 3-form RR field to which
the cylindrically wrapped brane couples. The only difference is dimensionality:
in case of the 2-brane the potential goes as $-R^2$, and for the 3-brane
it's $\sim - R^3$.} and goes like $\sim R^2\cdot D - R^3$ 
and is arbitrarily negative.
However, there is a potential energy barrier
and one needs to construct the bouncing euclidean solution in order to
address the problem in a complete way. In \cite{callan} it was attempted
to approach the problem by dropping the contributions due to the electric field, 
in that case the
lagrangian is Lorentz invariant with respect to r,t and it is possible to
construct some approximation to the bouncing solution.

 In this paper I will compute exactly the energy of the string and sphaleron
solution (the unstable static solution at the top of the potential barrier).
This will allow to conclude that the tunneling rate in fact goes to infinity
when the branes approach each other but still are at a finite distance
$\sim \sqrt{A}$.

\section{The Two Static Configurations}

I will review the construction of the releveant solution from \cite{callan}.
Similar solutions were also considered by Gibbons in \cite{gibbons}.

Consider the case when the worldbrane gauge field is purely electric
and only one transverse coordinate X is excited. The worldbrane action
reduces to
\be
L = -\frac{1}{g_p} \int d^{4}x \sqrt{ (1-\vec{E}^{2}) (1+ (\vec{\nabla} X)^{2})
+ (\vec{E} \vec{\nabla} X)^{2} - {\dot{X}}^{2}}~,
\label{langrangian}
\ee
where $g_p=(2\pi)^3g$, and $g$ is the string coupling ($\alpha^{\prime} =1$).

The canonical momentum associated with $\vec{A}$ is
\be
g_p \vec{\Pi}=  \frac{ \vec{E}(1+ (\vec{\nabla}X)^{2})-\vec{\nabla}X
(\vec{E} \vec{\nabla} X )  } { \sqrt{ (1-\vec{E}^{2})
(1+ (\vec{\nabla} X)^{2})
+ (\vec{E} \vec{\nabla}X)^{2} - {\dot{X}}^{2}} }~.
\label{momentum}
\ee
The constraint equation is $\vec{\nabla}\vec{\cdot\Pi}=0$. The Hamiltonian is

\be
H = \frac{1}{g_p} \sqrt{ (1+(g_p \vec{\Pi})^{2}) (1+ (\vec{\nabla} X)^{2})-
(g_p \vec{\Pi} \times \vec{\nabla}X)^2 }~~.
\label{ham}
\ee

We are looking for the most
general static, spherically symmetric solution. The equation for X, which
follows from varying the energy, after setting $\dot{X}=0$ is
\be
\vec{\nabla}  \frac{ (1-\vec{E}^{2}) \vec{\nabla}X +
\vec{E}(\vec{E} \vec{\nabla}X) }{ \sqrt{ (1-\vec{E}^{2})
(1+ (\vec{\nabla}X)^{2})
+ (\vec{E} \vec{\nabla }X)^{2} - {\dot{X}}^{2}} }  =0~.
\label{xeqn}
\ee

From (\ref{momentum}) follows that $g_p \vec{\Pi}= \frac{A \vec{r}}{r^3}$,
and from (\ref{xeqn})
$ \frac{\vec{\nabla} X}{\sqrt{1- E^2 + \nabla X^2 }}= \frac{B \vec{r}}{r^3}$.
Here A and B are integration constants.

Here
$$ g_p \vec{\Pi} = \frac{\vec{E}}{\sqrt{1-E^2 + \nabla X^2} }~.$$

The solution for the coordinate and electric field is
\be
\vec{\nabla} X=\frac{B\vec{r}}{\sqrt{r^4 + A^2 - B^2}}~~~~~~~~
\vec{E} = \frac{A\vec{r}}{\sqrt{r^4 + A^2 - B^2}}~.
\label{solution}
\ee

One can view this solution as a way to minimally break
supersymmetry, instead of $E=\nabla X$ we have $E=\frac{A}{B}\nabla X$.
In principle, A should be quantized as electric charge in units of $\pi g$.
We will be interested in $B > A$, in this case the resulting configuration
will be the 3-brane and anti-brane joined by a smooth throat. To see
this one needs to explicitly exploit the geometry by finding X:
\be
X(r)= B \int_{r}^{\infty} \frac{ds}{\sqrt{s^4 - r_0^4}}~. 
\label{shape}
\ee

Here $ r_0^4 = B^2 - A^2 $. We have set $X(\infty)=0$, e.g. far away
the brane is flat and is at zero coordinate in the transverse direction.
$X(r_0)$ is finite, but $X^{\prime}(r_0)$ is infinite: the throat becomes
vertical at that radius. This can be continued back out through $r_0$
to give the two branes. Branes possess orientation and continuing it through
the throat we see that the new brane is of the opposite orientation: 
an antibrane.

The relations between A,B and $r_0$, $X(r_0)=D/2$ can be reversed  
to express $r_0$ and B in terms of D and A:

\be
D/2= c \sqrt{\frac{A^2}{r_{0}^{2}} + r_{0}^{2}} ~~and~~~~~ B^2= A^2+ r_{0}^{4}~, 
~~~~~~~where ~~~ c = \int_1^{\infty} \frac{dz}{\sqrt{z^4-1}}~.
\label{dbeqn}
\ee
In the limit of large D
the two possible radii at the throat are $r_0 \sim D/2c~,$ and $~A~2c/D $.
A remark is in place here that the minimal separation for which a real root
exists is $D_{min} = 2c\sqrt{2A}$.

Knowing the energy function \\ $H = \sqrt{(1+ \nabla X^2)(1+ g\Pi^2)}$ allows
to compute the energies of the solutions exactly,

\be
E_{tot} = \frac{1}{g_p}\int_{r_0}^{\infty}
\sqrt{1+ \frac{A^2 + r_{0}^{4}}{r^4 - r_{0}^{4}}}~~
\sqrt{1+\frac{A^2}{r^4}} ~~4\pi r^2~ d r~~.
\label{energy}
\ee

At this point the temptation to make the problem completely dimensionless
becomes irresistible. Let me introduce the parameters 
$\mu=\frac{D}{2c\sqrt{A}},
~~~r=z\sqrt{A} $. I am now measuring length in units of $\sqrt{A}$, energy in
$A^{3/2}$, and  also $r_0 = \xi \sqrt{A}$:

\be
E_{1,2} = \frac{1}{g_p}\int_{\xi_{1,2}}^{\infty} \sqrt{1+ \frac{1 +
\xi^4}{z^4 - \xi^4}}~~
\sqrt{1 + \frac{1}{z^4}} ~~4\pi z^{2}~ dz~~.
\label{dimless}
\ee
One can check that $\xi_1 \cdot \xi_2 = 1$, because (\ref{dbeqn}) becomes
$\mu^2= \xi^2 + 1/\xi^2$. The minimum separation is $\mu_{min} = \sqrt{2}$,
at which point the two roots of the quadratic equation become degenerate:
$\xi_1 = \xi_2 = 1$.

After some manipulations with the energy integral, and a variable change \\
$y=\xi / z$, I get
\be
\frac{g_p}{8\pi}E = \xi^3 \int_{0}^{1} \frac{d y}{y^{4}\sqrt{1-y^4}} +
\frac{1}{\xi} \int_{0}^{1} \frac{d y}{\sqrt{1-y^4}}~~.
\label{answer}
\ee

 I have not dealt yet with the volume infinity which manifests
itself in the fact that the first of these integrals is divergent at 
$y \rightarrow 0$.
Regularize it by substracting $1/y^4$ from the integrand : if one were to
go back, it is exactly equivalent to computing the energy with respect
to the configuration when the two branes are flat, parallel and not joined
by any string. Denote the integrals in (\ref{answer}) as u and v, 

\begin{eqnarray}
u= \int_{0}^{1} \frac{d y}{y^4}(\frac{1}{\sqrt{1-y^4}}- 1) -
\int_{1}^{\infty} \frac{d y}{y^4}=
-\frac{1}{3} + \sum_{n=1}^{\infty} \frac{1}{4n-3}\frac{(2n-1)!!}{n!~ 2^n}
\nonumber\\
u= \frac{\sqrt{\pi}}{4} \frac{\Gamma(-3/4)}{\Gamma(-1/4}=.43701...
\nonumber\\
v=  \int_{0}^{1} \frac{dy}{\sqrt{1-y^4}}= \sqrt{\pi}
\frac{\Gamma(5/4)}{\Gamma(3/4)}=1.31103...
                        \label{numerics}
\end{eqnarray}
The answers are not surprising, since both integrals are generalized
B functions, and the second one is in fact the quarter period of the
Jacoby elliptic functions. One can also show, by using the functional
properties of the Gamma function, that $3u=v$. 
Also, constant c from (\ref{dbeqn})
is also related, in fact $c=v$. The energies of the string
and the sphaleron become, respectively
$$
E_{string}= u ( \xi^{-3} +3\xi)~~~~~ E_{sph}= u ( \xi^3 + 3\xi^{-1} )
$$
Note, that I have now taken the larger of the two roots to be $\xi$, the other
root being $1/\xi$. One can even write a relation between the energies of 
the string and sphaleron solutions  $E_{string}(1/\xi)=E_{sph}(\xi)$. 
Even though this relation is formal,
(each function is defined only for arguments larger than one) 
it might be of significance in the future. The energy of a long string
after fully substituting the dimensionful units becomes 
$$ E_{str}= \frac{4\pi}{g_p} D A = 
\frac{4\pi}{8\pi^3g}D \pi g= \frac{1}{2\pi}D$$.
\begin{figure}
\centerline{\hbox{\psfig{figure=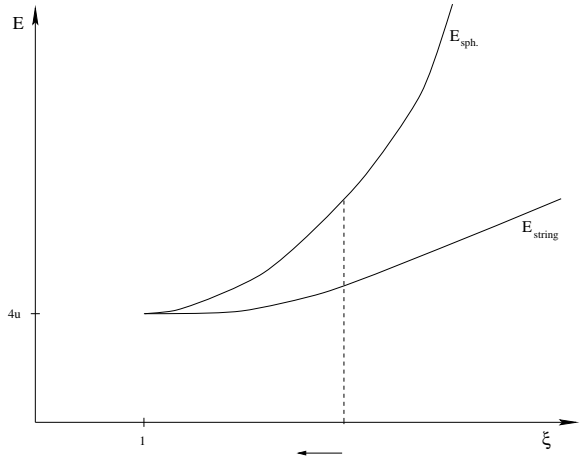,height=6cm,angle=0}}}
\caption[fig1]{Energy vs. $\xi$. \newline {\it ~~Height of the barrier}
 = $E_{sph}-E_{string} \rightarrow 0$    } 
\label{fig1}
\end{figure}
This reproduces the correct tension of the fundamental string.
The slope of the energy function is zero at
the minimum possible value of $\xi=1$ (Fig 1).

One  further interesting elaboration  is possible. The energy surface in the
parameter space of $r_0$ and D (or $\mu, \xi$) apparently has a kink well known from
catastrophe theory. The point at which the kink first appears corresponds to
$\mu^2=2$ and $\xi=1$. For $\mu^2 < 2$  the  energy is monotonous in $\xi$, and
there is no static solutions,
otherwise it has one minimum and one maximum, corresponding to the string
and the sphaleron respectively.  

\section{Annihilation by Tunneling.}

Let us recall that two branes, a 3-brane and an anti-brane, are going
to gravitationally attract, even though weakly, causing them to move closer and
to eventually annihilate. The tunneling through the potential barrier,
the sphaleron being the unstable solution at the top, will be
exponentially weak \cite{callan}, but as the branes move closer
D becomes smaller. So are $\mu$ and $\xi$. This process makes the tunneling
easier in a twofold way: first by making the barrier narrower (distance between
$\xi$ and $1/\xi$ becomes smaller), and second, the energy at the top decreasing
to become equal to the energy of the string.
As $\xi \rightarrow 1$, the rate of tunneling rate increases indefinitely,
and at $\xi=1$ the metastable string state ceases to exist. This is an
interesting example of physical continuity: before the state disappeared as we
change the parameter (D or $\xi$), its physical width due to tunneling had to become
infinite. 

 Unfortunately, even though these conclusions are self consistent, 
they may still be physically
incomplete due to the fact that as branes approach to planckian distance
new nonperturbative phenomena of annihilation may kick in. One possible
way out of this difficulty is to crank up the string coupling, this leads into
unknown territory too but our quasi-classical tunneling 
may indeed be the dominant mode of annihilation in that regime.

\section{Acknowledgements}
This work was finished in December 1997, I decided that these calculationd might be
interesting after similar formulas appeared in \cite{juan} and \cite{gross} 
in calculations of the quark-anti-quark potential.
I am grateful to Professor C.Callan for introducing me to this area of research
and in fact suggesting this problem.
I would also like to thank W.Taylor, G.Savvidy and S.Lee for useful discussions.

\vfill

\begin{thebibliography}{99}

\bibitem{callan} 
C.G.~Callan and J.M.~Maldacena, ``Brane Dynamics from the Born-Infeld Action,'' 
Nucl.~Phys. {\bf B513} (1998) 198, 
{{\tt  hep-th/9708147}}.
\bibitem{emparan} R.~Emparan, ``Born-Infeld Strings Tunneling to D-branes,''
{{\tt hep-th/9711106 }}
\bibitem{gibbons} G.W.~Gibbons, ``Born-Infeld particles and Dirichlet p-branes,''
Nucl.~Phys. {\bf B514} (1998) 603, 
{{\tt  hep-th/9709027}}.
\bibitem{juan} J.~Maldacena, ``Wilson loops in large N field theories,''
{{\tt hep-th/9803002 }}
\bibitem{gross} D.J.~Gross, H.~Ooguri, ``Aspects of large N gauge theory dynamics as
seen by string theory,'' 
{{\tt hep-th/9805129}}.
\end{thebibliography}
\end{document}